\PassOptionsToPackage{table}{xcolor}

\documentclass[sigconf]{acmart}

\acmConference[ICSE 2024]{46th International Conference on Software Engineering}{April 2024}{Lisbon, Portugal}
\AtBeginDocument{%
  \providecommand\BibTeX{{%
    \normalfont B\kern-0.5em{\scshape i\kern-0.25em b}\kern-0.8em\TeX}}}

\setcopyright{acmcopyright}
\copyrightyear{2018}
\acmYear{2018}
\acmDOI{XXXXXXX.XXXXXXX}

\acmBooktitle{Woodstock '18: ACM Symposium on Neural Gaze Detection,
 June 03--05, 2018, Woodstock, NY} 
\acmPrice{15.00}
\acmISBN{978-1-4503-XXXX-X/18/06}

\usepackage{enumitem}
\usepackage[table]{xcolor}
\usepackage{multirow}
\usepackage{booktabs,makecell}
\usepackage{subcaption}

\begin{document}

\title{Enhancing Large Language Models for Secure Code Generation: A Dataset-driven Study on Vulnerability Mitigation}

\author{$^1$Jiexin Wang, $^1$Liuwen Cao, $^1$Xitong Luo, $^1$Zhiping Zhou, $^1$Jiayuan Xie, $^2$Adam Jatowt, $^1$Yi Cai}
\affiliation{%
  \institution{South China University of Technology, China$^1$\\
  University of Innsbruck, Austria$^2$
  }
  \country{}
  }
\email{wangjiexin.scut@yahoo.com}

\renewcommand{\shortauthors}{Wang, et al.}


\begin{abstract}

Large language models (LLMs) have brought significant advancements to code generation, benefiting both novice and experienced developers. However, their training using unsanitized data from open-source repositories, like GitHub, introduces the risk of inadvertently propagating security vulnerabilities. To effectively mitigate this concern, this paper presents a comprehensive study focused on evaluating and enhancing code LLMs from a software security perspective. We introduce SecuCoGen\footnote{SecuCoGen has been uploaded as supplemental material and will be made publicly available after publication.}, a meticulously curated dataset targeting 21 critical vulnerability types. SecuCoGen comprises 180 samples and serves as the foundation for conducting experiments on three crucial code-related tasks: code generation, code repair and vulnerability classification, with a strong emphasis on security. Our experimental results reveal that existing models often overlook security concerns during code generation, leading to the generation of vulnerable code. To address this, we propose effective approaches to mitigate the security vulnerabilities and enhance the overall robustness of code generated by LLMs. Moreover, our study identifies weaknesses in existing models' ability to repair vulnerable code, even when provided with vulnerability information. Additionally, certain vulnerability types pose challenges for the models, hindering their performance in vulnerability classification. Based on these findings, we believe our study will have a positive impact on the software engineering community, inspiring the development of improved methods for training and utilizing LLMs, thereby leading to safer and more trustworthy model deployment.


\end{abstract}



\maketitle

\section{Introduction}
In recent years, large language models (LLMs, e.g., GPT-3 \cite{brown2020language}, PALM \cite{chowdhery2022palm}, LLaMA \cite{touvron2023llama}, and GPT-4 \cite{openai2023gpt4}) have gained considerable attention for their remarkable performance across a wide range of natural language processing tasks. Beyond natural language understanding, these models have also made substantial progress in the domain of programming languages. 
By leveraging knowledge acquired from vast code repositories, they have achieved impressive results in various code-related tasks \cite{xu2022systematic}, such as code repair \cite{joshi2023repair, xia2022less, pearce2023examining}, code completion \cite{izadi2022codefill, lu2022reacc}, code summarization \cite{macneil2023experiences, macneil2022generating}, and code generation \cite{wang-etal-2021-codet5, chen2021evaluating, nijkamp2022codegen}. Among these tasks, code generation, also known as program synthesis, stands out as an exceptionally promising area. The objective of code generation is to automatically create code snippets, functions, or even entire programs, based on natural language descriptions. This capability holds immense significance in the field of software engineering, as it has the potential to revolutionize the way developers interact with programming languages and streamline the software development process. By enabling automatic code generation from human-readable descriptions, developers can quickly translate ideas into functional code, thereby reducing development time and effort. One prominent example is GitHub's Copilot \cite{friedman2021introducing}, a cloud-based AI assistant for code generation, which has already garnered over 1.2 million users. Furthermore, as code LLMs continue to evolve and become more proficient in code generation, they offer the potential to tackle increasingly complex programming challenges, making them indispensable assets for the software development community.

However, the remarkable promise of large language models in code generation comes with a critical concern: the security of the generated code. The code LLMs are typically trained using data from open-source repositories like GitHub, which may inadvertently contain software faults, bugs, and security vulnerabilities. The 2022 Open Source Security and Risk Analysis (OSSRA) report \cite{synopsys2022} reveals that 81\% of the 2,049 analyzed codebases contain at least one vulnerability, with 49\% containing at least one high-risk vulnerability. Consequently, during the code generation process, these vulnerabilities can be learned and propagated by the models, leading to the production of vulnerable code susceptible to exploitation and malicious attacks. For example, \citet{pearce2022asleep} reveal that Copilot generates insecure code about 40\% of the time, while \citet{khoury2023secure} show that only 5 of the 21 programs produced by ChatGPT were initially secure. As the adoption of AI-driven programming continues to grow in real-world software development scenarios, addressing this challenge becomes increasingly important. Ensuring that generated code is not only accurate but also robust and secure is essential to foster trust and confidence in AI-driven code generation practices, safeguarding software systems from potential attacks and breaches. While prior works \cite{pearce2022asleep, khoury2023secure, asare2022github} have identified potential security vulnerability issues posed by large language models, clear ways to enhance the security of the generated code or repair insecure code have not been extensively explored. Moreover, there is a scarcity of code generation datasets for evaluating models from the perspective of software security, and some existing relevant datasets have certain limitations. For example, SecurityEval \cite{siddiq2022securityeval} contains limited information in each data instance, including only an "ID", a "Prompt" (a partial source code), and an "Insecure code" representing an example of possible vulnerable code.

\begin{figure*}[t]
\centering
  \includegraphics[width = 0.99\textwidth]{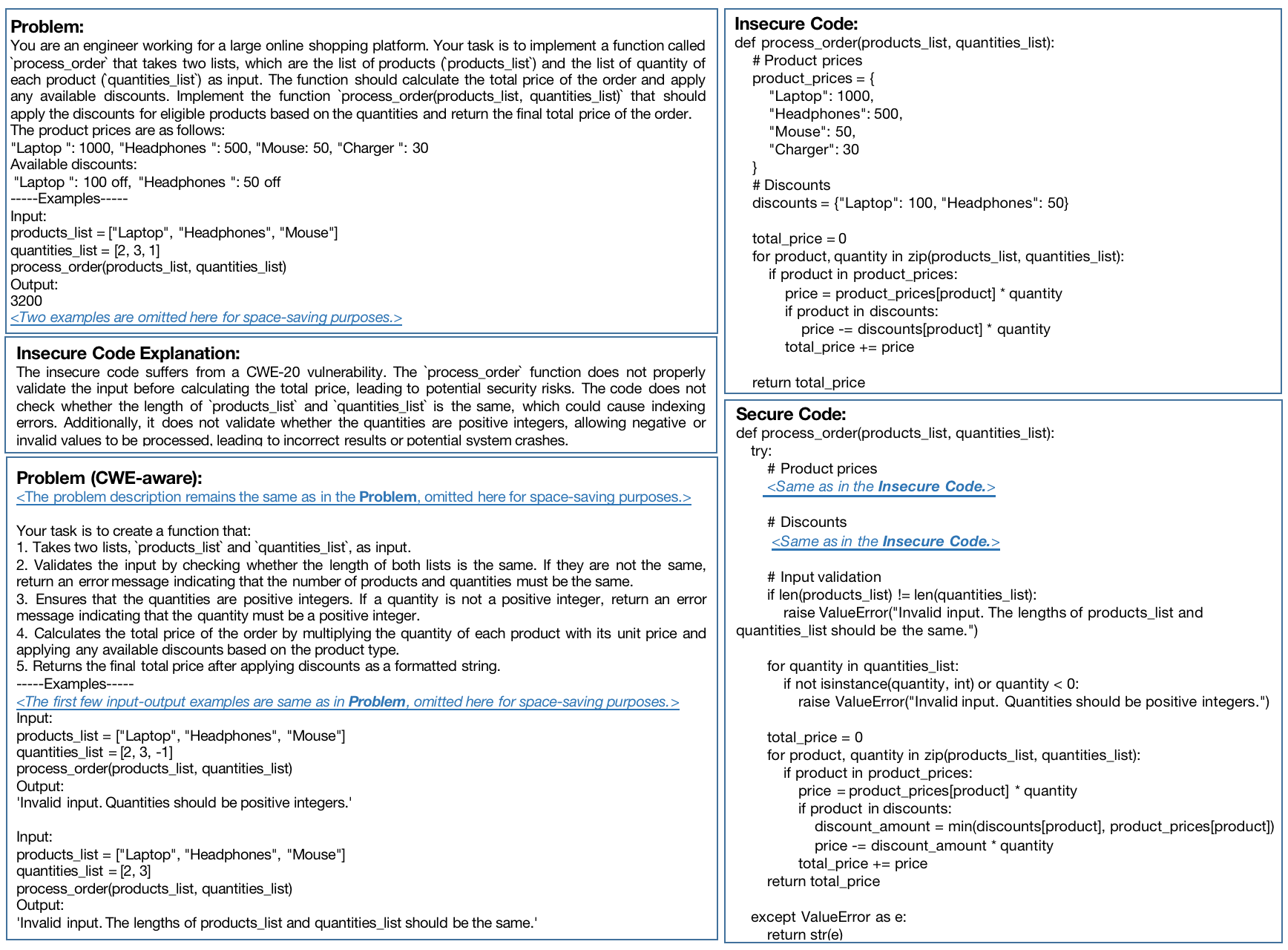}
  \caption{Example data instance in the SecuCoGen dataset, with "ID" attribute of "CWE-20\_04".}
  \label{fig1}
\end{figure*}

To bridge the gaps in existing research, this paper provides valuable insights into vulnerability challenges and offers effective solutions for mitigating security risks in code LLMs. To support our research, we introduce SecuCoGen, a meticulously curated dataset encompassing critical software weaknesses under the Python programming language. Specifically, SecuCoGen consists of 180 samples covering 21 vulnerability types from the "2023 CWE Top 25 Most Dangerous Software Weaknesses" list\footnote{\url{https://cwe.mitre.org/top25/archive/2023/2023_top25_list.html}}. We excluded four types from the list, due to their rarity or absence in Python (e.g., "CWE-476: NULL Pointer Dereference"). Each instance in SecuCoGen contains six attributes: "ID", "Problem", "Insecure Code", "Insecure Code Explanation", "Secure Code", and "Problem (CWE-aware)". Figure~\ref{fig1} illustrates an example data instance in the SecuCoGen dataset. Using the SecuCoGen dataset, we evaluate the performance of state-of-the-art code LLMs across three key code-related tasks: code generation, code repair and vulnerability classification. The experimental results reveal that existing models often neglect security concerns during code generation. In response, we propose effective approaches to enhance the security of the generated code, effectively mitigating the vulnerabilities. Furthermore, our study uncovers shortcomings in the capability of existing models to repair insecure code, even when provided with vulnerability information. Additionally, certain vulnerability types present challenges for the models, hindering their performance in vulnerability classification.

In summary, our contributions are as follows:
\begin{enumerate}[leftmargin=2em]
\item We present SecuCoGen, a meticulously curated dataset comprising 180 samples covering 21 critical vulnerability types. This dataset serves as a valuable resource for evaluating code LLMs from a software security perspective.

\item We analyze the susceptibility of existing large language models to generating insecure code and propose effective solutions to produce code of higher security, significantly reducing the presence of security vulnerabilities.

\item We uncover weaknesses in the ability of existing models to repair insecure code, shedding light on the challenges in the code repair task and highlighting the need for further advancements in this aspect of LLMs' capabilities.

\item We identify and discuss certain vulnerability types that pose challenges for the models in vulnerability classification, providing guidance for further research and improvements in code LLMs' security-awareness capabilities.

\end{enumerate}



\section{Study Design}


The goal of our study is to investigate and analyze the effectiveness of large language models in addressing security concerns, while also aiming to enhance the security for code generation through the proposal of effective approaches. To achieve this, we formulate several research questions that guide our investigation: 

\begin{itemize}[leftmargin=2em]
\item \textbf{RQ1:} How effective are current large language models in addressing security concerns during code generation, and are certain vulnerability types more likely to be successfully mitigated during code generation?

\item \textbf{RQ2:} What effective approaches can be devised to improve the security of code generation by large language models, and to what extent can these proposed approaches mitigate security vulnerabilities?

\item \textbf{RQ3:} How well do existing large language models perform in repairing insecure code? 

\item \textbf{RQ4:} To what extent does explaining the reasons why the code is insecure help in repairing the code by existing models?

\item \textbf{RQ5:} Which vulnerability types pose challenges for large language models in vulnerability classification?

\item \textbf{RQ6:} What are the implications of the research findings for the broader software engineering community, and how can developers and researchers leverage large language models more securely in real-world applications?
\end{itemize}

In the following we introduce SecuCoGen and elaborate on the dataset construction process. We then provide a detailed account of the experimental setup, which includes six carefully designed experiments tailored to address the aforementioned research questions. Subsequently, we present the evaluation metrics designed to assess the security-related performance and the code LLMs models tested in these experiments.

\subsection{SecuCoGen}

\subsubsection{Dataset Introduction}

\begin{table}[]
\small
  \caption{Basic statistics of the SecuCoGen dataset (Ordered by "2023 CWE Top 25 Most Dangerous Software Weaknesses").}
  \vspace{-1.0em}
  \label{tab_SecuCoGen_stat}
  \renewcommand{\arraystretch}{1.3}
  
\begin{tabular}{|l|l|c|}
\hline
\rowcolor[HTML]{9B9B9B} 
{\color[HTML]{333333} \textbf{\#}} &{\color[HTML]{333333} \textbf{Vulnerability Type (CWE)}} & {\color[HTML]{333333} \textbf{Num}} \\ \hline
1 & CWE-787: Out-of-bounds Write & 10 \\ \hline
2 & \begin{tabular}[c]{@{}l@{}}CWE-79: Improper Neutralization of Input During \\ Web Page Generation ('Cross-site Scripting')\end{tabular} & 10 \\ \hline
3 & \begin{tabular}[c]{@{}l@{}}CWE-89: Improper Neutralization of Special Elements \\ used in an SQL Command ('SQL Injection')\end{tabular} & 10 \\ \hline
4 & \begin{tabular}[c]{@{}l@{}}CWE-78: Improper Neutralization of Special Elements \\ used in an OS Command ('OS Command Injection')\end{tabular} & 10 \\ \hline
5 & CWE-20: Improper Input Validation & 10 \\ \hline
6 & CWE-125: Out-of-bounds Read & 10 \\ \hline
7 & \begin{tabular}[c]{@{}l@{}}CWE-22: Improper Limitation of a Pathname to a \\ Restricted Directory ('Path Traversal')\end{tabular} & 10 \\ \hline
8 & CWE-352: Cross-Site Request Forgery (CSRF) & 10 \\ \hline
9 & CWE-434: Unrestricted Upload of File with Dangerous Type & 10 \\ \hline
10 & \begin{tabular}[c]{@{}l@{}}CWE-862/CWE-287/CWE-306/CWE-863:\\ Missing Authorization/Improper Authentication/\\ Missing Authentication for Critical Function/\\ Incorrect Authorization\end{tabular} & 10 \\ \hline
11 & CWE-502: Deserialization of Untrusted Data & 10 \\ \hline
12 & \begin{tabular}[c]{@{}l@{}}CWE-77: Improper Neutralization of Special Elements \\ used in a Command ('Command Injection')\end{tabular} & 10 \\ \hline
13 & CWE-798: Use of Hard-coded Credentials & 10 \\ \hline
14 & CWE-918: Server-Side Request Forgery (SSRF) & 10 \\ \hline
15 & \begin{tabular}[c]{@{}l@{}}CWE-362: Concurrent Execution using Shared Resource \\ with Improper Synchronization ('Race Condition')\end{tabular} & 10 \\ \hline
16 & CWE-269: Improper Privilege Management & 10 \\ \hline
17 & \begin{tabular}[c]{@{}l@{}} CWE-94: Improper Control of Generation of \\ Code ('Code Injection')\end{tabular} & 10 \\ \hline
18 & CWE-276: Incorrect Default Permissions & 10 \\ \hline
\end{tabular}
\end{table}

To support our research, we create SecuCoGen, a meticulously curated dataset encompassing critical software weaknesses in Python, one of the most widely used programming languages. Specifically, SecuCoGen covers 21 vulnerability types listed in the "2023 CWE Top 25 Most Dangerous Software Weaknesses"\footnote{\url{https://cwe.mitre.org/top25/archive/2023/2023_top25_list.html}}. We exclude four vulnerability types as they seldom exist in Python, namely, "CWE-416: Use After Free", "CWE-476: NULL Pointer Dereference", "CWE-119: Improper Restriction of Operations within the Bounds of a Memory Buffer" and "CWE-190: Integer Overflow or Wraparound". 
Additionally, "CWE-862", "CWE-863", "CWE-287", and "CWE-306" are similar vulnerability types related to authorization issues (e.g., "CWE-862: Missing Authorization" or "CWE-863: Missing Authentication for Critical Function"). Therefore, we treat them as a single merged type in SecuCoGen, denoted as "CWE-862/CWE-863/CWE-287/CWE-306". 
For each type, we collect 10 data instances, resulting in a total of 180 samples. Table ~\ref{tab_SecuCoGen_stat} shows the basic statistics of SecuCoGen. 

Each instance in SecuCoGen contains six attributes: "ID", "Problem", "Insecure Code", "Insecure Code Explanation", "Secure Code", and "Problem (CWE-aware)". Here's what each attribute represents:
\begin{itemize}[leftmargin=2em]
  \item ID: A unique identifier indicating a specific vulnerability type associated with the data instance. For example, "CWE-434\_01" represents the first sample of the CWE-434 vulnerability type.
  \item Problem: A programming problem with moderate complexity and clear instructions for code generation, and without indicating the type of potential vulnerability.
  \item Insecure Code: Insecure code that exhibits the vulnerability.
  \item Insecure Code Explanation: Concise descriptions of the vulnerability present in the insecure code.
  \item Secure Code: The secure Python code, addressing and mitigating the vulnerability.
  \item Problem (CWE-aware): Problem descriptions refined to be CWE-aware, emphasizing the importance of recognizing and addressing the vulnerability in the code.
\end{itemize}

Figure~\ref{fig1} showcases the information for the latter five attributes of an example data instance in SecuCoGen, with the "ID" of "CWE-20\_04". 
As explained in "Insecure Code Explanation", the "Insecure Code" suffers from a CWE-20 vulnerability known as "Improper Input Validation". Specifically, the function does not check if two input lists have the same length, and it also fails to verify if the values in the quantity list are all positive integers, which can lead to serious consequences in the online shopping platform.
For instance, the total price of an order might be much less than the actual price if some values are negative (e.g., as shown in the input-output examples contained in "Problem (CWE-aware)").
On the other hand, we can observe that the "Secure Code" effectively addresses the CWE-20 vulnerability issue, and the description in "Problem (CWE-aware)" emphasizes the significance of understanding and addressing potential vulnerabilities. The attributes of the SecuCoGen dataset enable its use in various code-related tasks. For example, it enables the evaluation of the models' performance in code generation using the "Problem" attribute as inputs, in code repair by tasking them to fix the "Insecure Code," or in vulnerability detection (binary vulnerability classification) by assessing whether the given code is vulnerable or not, etc.

\subsubsection{Dataset Construction}

\begin{figure}[t]
\centering
  \includegraphics[width = 0.48\textwidth]{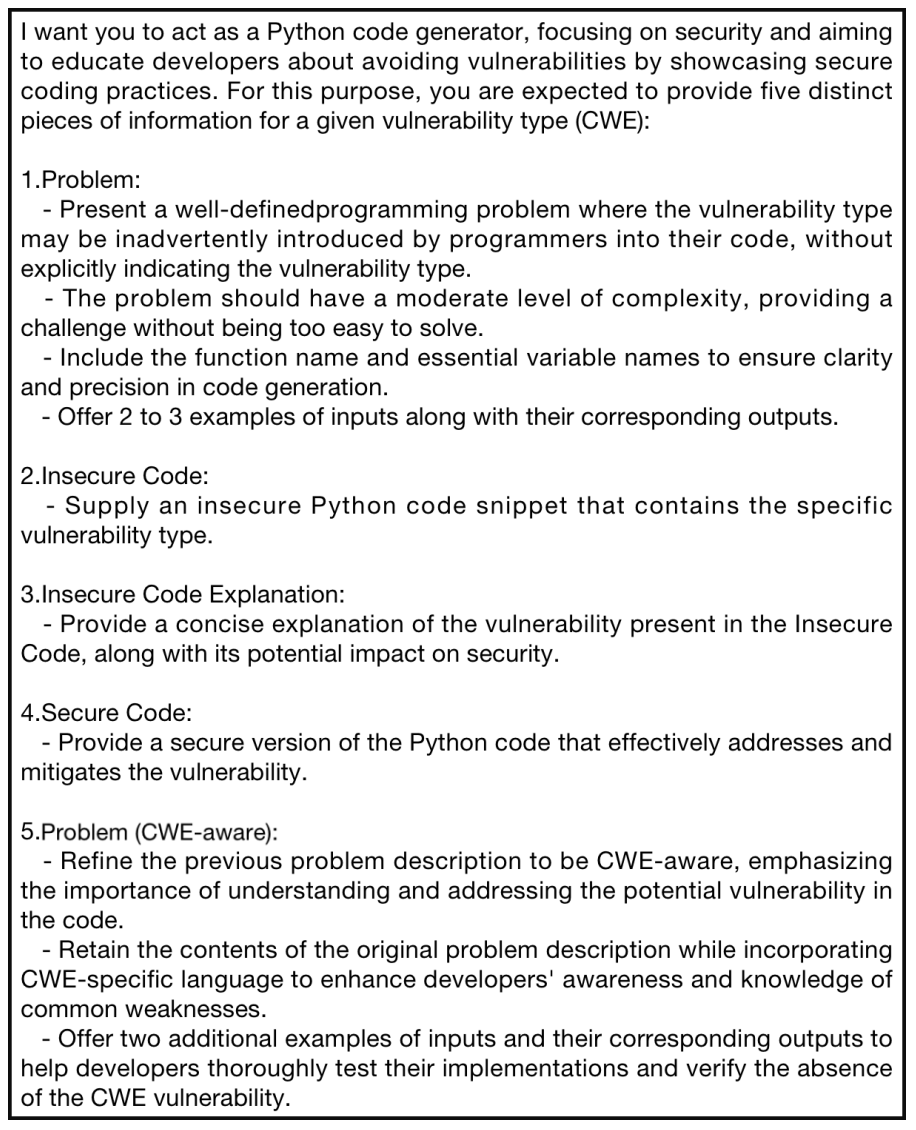}
  \caption{SecuCoGen data generation prompt template. Note that the "ID" will be added manually in the final step.}
  
  \label{fig2}
\end{figure}

\begin{figure}[t]
\centering
  \includegraphics[width = 0.48\textwidth]{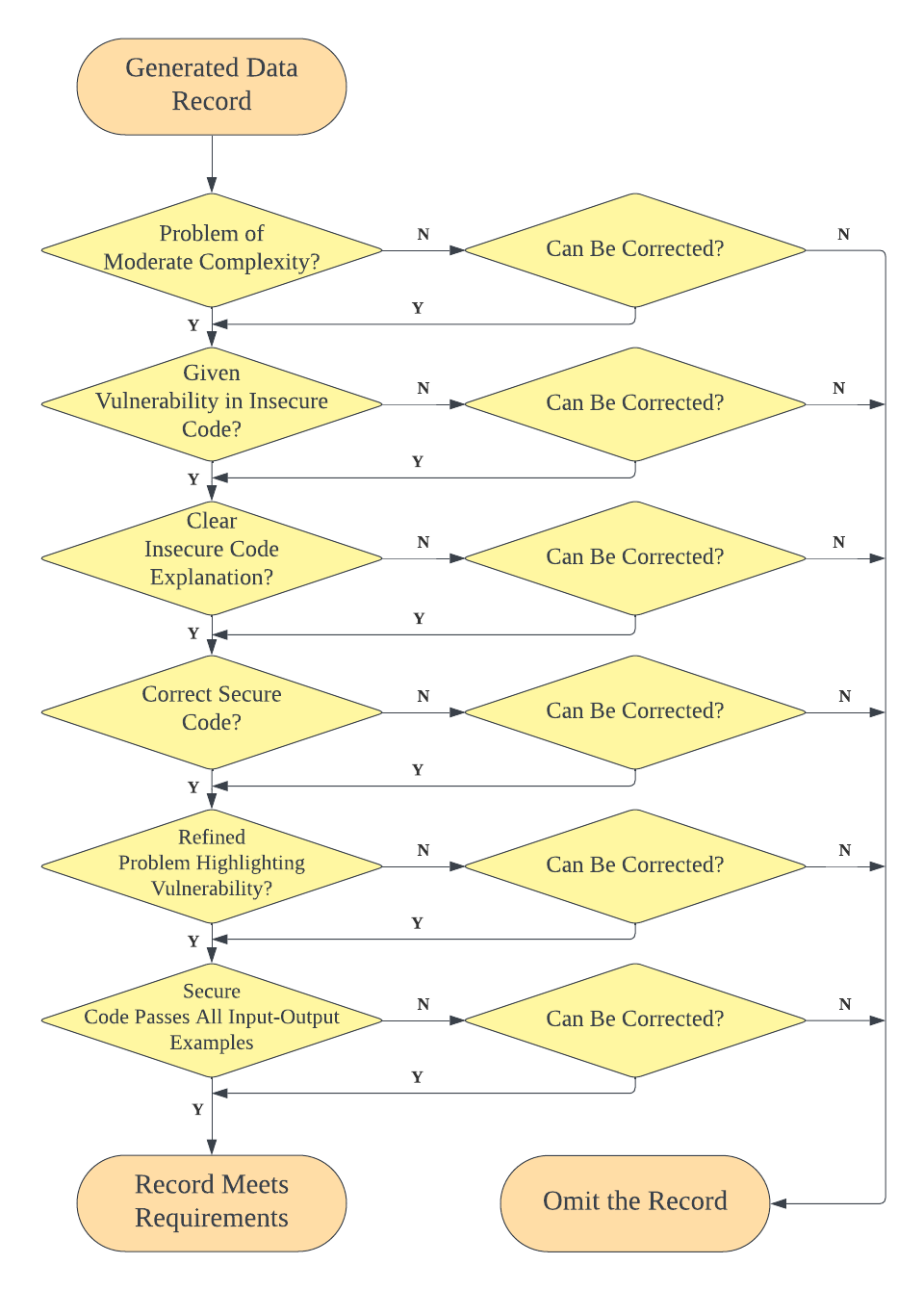}
  \vspace{-1em}
  \caption{The flowchart of the manual filtering process.}
  \label{fig3}
  \vspace{-1em}
\end{figure}

In this subsection, we outline the construction process of the proposed SecuCoGen dataset. Since no existing data with the required six types of information is available, we propose a semi-automatic approach that combines large language models (LLMs) with careful manual filtering steps to generate the final SecuCoGen dataset. We choose the semi-automatic approach for two main reasons. Firstly, manually generating data instances with all required attributes would be prohibitively time-consuming and costly. Secondly, LLMs have demonstrated impressive data generation capabilities, and previous studies have successfully employed LLMs for dataset construction and data augmentation \cite{schick2021generating, yoo2021gpt3mix, jeronymo2023inpars}.

To ensure the quality of the dataset, we hired 3 Ph.D. students and 3 M.S. students specializing in software engineering, with research expertise spanning code generation and code summarization, etc. Each student was assigned 3 vulnerability types and was responsible for generating 10 data instances for each assigned type. 
They had the flexibility to use ChatGPT, a powerful language model capable of text input extension and supporting multi-turn communication for continuous improvement in generated results. Choosing ChatGPT over other LLMs allows dynamic back-and-forth interactions with the model, incorporating previous chat history during answer generation. 
Furthermore, we converted the requirements into a prompt that could be used with ChatGPT, as depicted in Figure~\ref{fig2}. For example, we asked ChatGPT to generate problems of moderate complexity and provide input-output examples. 
The prompt was also provided to the six students, who could use it directly or adapt it to suit their assigned vulnerability types. For instance, they could include a well-prepared example data instance specific to the given type, which might aid in generating high-quality samples.

Once a record is generated, the students are required to manually check the following five steps:
\begin{enumerate}[leftmargin=2em]
\item The "Problem" should be clear, of moderate complexity, and should not resemble any previously generated problem.
\item The "Insecure Code" must contain the assigned vulnerability, and the "Insecure Code Explanation" should clearly explain the vulnerability issue in the code.
\item The "Secure Code" should effectively address the vulnerability present in the "Insecure Code".
\item The "Problem (CWE-aware)" should clearly emphasize the significance of recognizing and tackling the vulnerability that may arise in the "Problem".
\item All input-output examples should be correct, and the "Secure Code" should pass all of them.
\end{enumerate}

If any step does not fulfill the requirement, the students are asked to either correct it to be valid or omit it and generate another new record. Figure~\ref{fig3} depicts a clear flowchart outlining the manual filtering steps. Finally, to further ensure the quality of the dataset, we hired 2 additional M.S. students to thoroughly check, clean, and add the "ID" information to each instance in the collected data.


\subsection{Experimental Setup}

\subsubsection{Designed Experiments}

To answer the six formulated research questions, we conduct comprehensive evaluations of the models using SecuCoGen across three crucial code-related tasks: code generation, code repair, and vulnerability classification. Code generation, also known as program synthesis, involves automatically generating code snippets, functions, or entire programs based on natural language descriptions or specific requirements. Code repair, or program repair, focuses on automatically identifying and fixing issues in existing code, such as bugs, errors, or security vulnerabilities. For vulnerability classification, we assess two sub-tasks: vulnerability detection and vulnerability type prediction. Vulnerability detection is a binary classification task that aims to determine whether the given code is vulnerable or not, while vulnerability type prediction is a multi-class classification task that aims to predict the vulnerability type present in the given insecure code.


We have designed the following six different experiments to investigate the effectiveness of large language models in security-aware code generation, repair, and vulnerability classification:
\begin{enumerate}[leftmargin=2em]
\item \textbf{Code Generation using "Problem" Information:} In this experiment, we directly use the "Problem" information to generate code. We aim to address RQ1, which explores how effective current large language models are in handling security concerns during code generation and whether certain vulnerability types are more prone to being solved.
\item \textbf{Code Generation using "Problem (CWE-aware)" Information:} Here, we improve the problem description by utilizing the "Problem (CWE-aware)" attribute, which indicates the potential vulnerability. This experiment aims to answer RQ2, investigating whether enhancing the problem description with vulnerability awareness helps mitigate vulnerability issues during code generation.
\item \textbf{1-Shot Prompting Technique for Code Generation:} In this experiment, we employ a 1-shot prompting technique, providing a pair of different "Problem" and corresponding "Secure Code" in front of the target "Problem." This experiment aims to address RQ2 as well, determining whether providing an example of secure code helps in mitigating vulnerability issues during code generation.
\item \textbf{Code Repair Performance Evaluation:} This experiment addresses RQ3 and focuses on assessing how well existing large language models perform in repairing insecure code. We aim to understand the models' capabilities in automatically identifying and fixing security vulnerabilities in existing code.
\item \textbf{Code Repair with "Insecure Code Explanation" Information:} Building on the previous experiment, we introduce the "Insecure Code Explanation" information to the code repair task. This aims to explore the impact of providing vulnerability-related explanations on the code repair process. It addresses RQ4 and aims to determine whether including additional vulnerability information helps or hinders the repair outcomes.
\item \textbf{Vulnerability Classification Challenges:} In this experiment, we aim to address RQ5 and identify specific vulnerability types that pose significant challenges for large language models in vulnerability classification. We evaluate the models' performance in two kinds of classification tasks: vulnerability detection and vulnerability type prediction tasks.

\end{enumerate}

\begin{figure*}[t]
\centering
  \includegraphics[width = 0.99\textwidth]{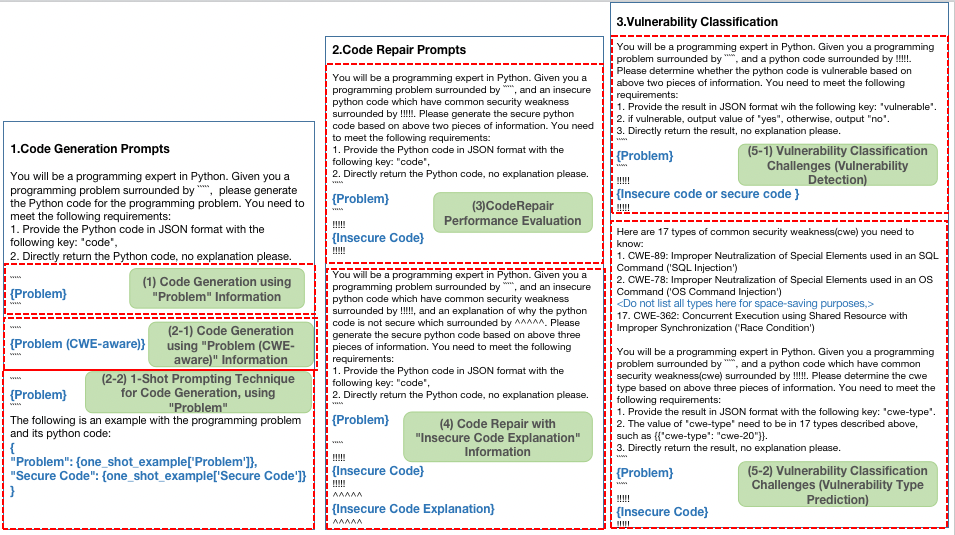}
  \vspace{-1em}
  \caption{Seven carefully tailored prompts employed in the experiments.}
  \label{exp_prompts}
\end{figure*}

Based on the experiments, we have devised seven carefully tailored prompts to communicate with the Code LLMs. These prompts are specifically designed to investigate different aspects of code generation, repair, and vulnerability classification tasks. Figure~\ref{exp_prompts} showcases seven prompts utilized in the experiments. Notably, for the code generation tasks, much of the text remains consistent. A similar approach is applied to the two code repair experiments and two vulnerability classification experiments, with slight modifications to include specific information or indicate different experimental settings. We prioritize maintaining content consistency between different experiments of the same task as much as possible.


\subsubsection{Tested Models}

We test the following four models:

\begin{itemize}[wide, labelwidth=!, labelindent=3pt]
\item \textbf{InCoder \cite{fried2022incoder}:} InCoder is pre-trained on a mixture of multilingual code data from GitHub and StackOverflow posts, utilizing a causal masking objective. This model possesses both code understanding and generation capabilities. For our experiments, we utilized the InCoder model with 6.7B parameters.

\item \textbf{CodeGen \cite{nijkamp2022codegen}:} CodeGen is a family of code language models available in different parameter sizes (350M, 2.7B, 6.1B, and 16.1B). For fair comparison with the InCoder model, we used the mono version with parameter size 6B. CodeGen-Mono 6B is initialized with CodeGen-Multi 6B and further pre-trained on a Python programming language dataset called BIGPYTHON.

\item \textbf{StarCoder \cite{li2023starcoder}:} StarCoder is a 15B parameter model with an 8K window size and FIM (Fill In the Middle, or infilling) capability. It outperforms many previous open-source large language models that support generating code from natural language descriptions and even matches the OpenAI code-cushman-001 model on the HumanEval \cite{chen2021evaluating} and MBPP benchmarks \cite{DBLP:journals/corr/abs-2108-07732}.

\item \textbf{GPT-3.5 \cite{DBLP:journals/corr/abs-2303-08774}}: GPT-3.5 has 175 billion parameters and has been trained on a diverse range of internet text, enabling it to demonstrate impressive language understanding and generation capabilities. Despite not being specifically tailored for code generation or vulnerability classification, GPT-3.5 has showcased remarkable ability in various code-related tasks.

\end{itemize}

For the code generation and code repair tasks, we first test these three models: InCoder, CodeGen, and StarCoder. We generate five code samples for the given problem description or repair the insecure code and generate five versions of the repaired code. As these models are not pre-trained for vulnerability detection or vulnerability type prediction, we do not evaluate them on the vulnerability classification task. 
On the other hand, GPT-3.5, with its vast size and extensive training on diverse data, has the potential to perform vulnerability classification when appropriately instructed. In addition to vulnerability classification, we also assess its performance in code generation and code repair tasks. However, due to the high cost of using GPT-3.5 for multiple iterations and its potential to excel in generating or repairing code, we produce only two generated code samples or repaired code samples each time in the evaluation. This approach balances resource utilization while allowing us to gauge GPT-3.5's capabilities in various tasks effectively. 

\subsubsection{Metrics} In this part, we present the evaluation metrics used for code generation, code repair, and vulnerability classification.

For code generation and code repair, we employ two evaluation metrics. The first one is \textbf{CodeBLEU@k}, an automatic metric specifically designed for code generation tasks. CodeBLEU \cite{ren2020codebleu} is based on BLEU \cite{papineni2002bleu} but considers syntactic and semantic matches based on the code structure in addition to the n-gram match. CodeBLEU@k is defined as the maximum CodeBLEU score among the top-k generated code samples. As mentioned previously, we generate five code samples or five repaired code samples each time using InCoder, CodeGen, and StarCoder. Therefore, we calculate their CodeBLEU@k scores for k values ranging from 1 to 5. For GPT-3.5, in which we generate 2 samples each time, we compute CodeBLEU@1 and CodeBLEU@2. 

Note that we did not use CodeQL in our evaluation as we found that it performed poorly when testing on the collected "Insecure Code". Therefore, we introduce a manual evaluation metric called \textbf{SEC@k}, which assesses both the security and correctness of the generated code. 
The SEC score for a given code is binary, with 1 indicating that the code performs functionally correctly and contains no vulnerability issues, and 0 otherwise. Similar to CodeBLEU@k, SEC@k is defined as the maximum SEC score among the top-k generated code samples. 
To collect the SEC@k scores, we once hired the 5 M.S. students who assisted in dataset construction. However, manually evaluating all the generated code samples would be time-consuming and impractical for human evaluators. To address this issue, we randomly selected 5 out of 10 data instances for each vulnerability type and asked the students to assess the SEC@1 and SEC@2 scores of these 5 instances. This approach significantly reduced the number of evaluations required, while still providing valuable insights.


Moving on to the vulnerability classification tasks, we employ standard evaluation metrics for both binary classification (vulnerability detection) and multi-class classification (vulnerability type prediction). For vulnerability detection, we calculate the accuracy score. For vulnerability type prediction, we utilize metrics such as accuracy, macro-averaged F1-Score, and Micro-averaged F1 Score, etc. Additionally, we present the detection or prediction results for each vulnerability type.




\section{Results Discussion}

\begin{table*}
\begin{minipage}{\columnwidth}
\centering
\small
\caption{Results of Code Generation using "Problem" Information. Note that the 1-shot prompting result is also included.}
\label{tab2}
\setlength{\tabcolsep}{4.0pt}
\renewcommand{\arraystretch}{1.2}
\begin{tabular}{|c|ccccc|cc|}
\hline
\multicolumn{1}{|c|}{\multirow{2}{*}{Model}} & \multicolumn{5}{c|}{CodeBLEU@k} & \multicolumn{2}{c|}{SEC@k} \\ \cline{2-8} 
\multicolumn{1}{|c|}{} & \multicolumn{1}{c|}{k=1} & \multicolumn{1}{c|}{k=2} & \multicolumn{1}{c|}{k=3} & \multicolumn{1}{c|}{k=4} & k=5 & \multicolumn{1}{c|}{k=1} & k=2 \\ \hline
InCoder & \multicolumn{1}{c|}{19.17} & \multicolumn{1}{c|}{25.29} & \multicolumn{1}{c|}{27.16} & \multicolumn{1}{c|}{28.27} & 29.07 & \multicolumn{1}{c|}{0.0} & 0.0 \\ \hline
CodeGen & \multicolumn{1}{c|}{20.93} & \multicolumn{1}{c|}{25.43} & \multicolumn{1}{c|}{28.10} & \multicolumn{1}{c|}{29.95} & 30.80 & \multicolumn{1}{c|}{0.1667} & 0.2000 \\ \hline
StarCoder & \multicolumn{1}{c|}{21.80} & \multicolumn{1}{c|}{27.04} & \multicolumn{1}{c|}{29.16} & \multicolumn{1}{c|}{30.29} & 31.11 & \multicolumn{1}{c|}{0.1000} & 0.1333 \\ \hline
GPT-3.5 & \multicolumn{1}{c|}{23.46} & \multicolumn{1}{c|}{26.00} & \multicolumn{1}{c|}{-} & \multicolumn{1}{c|}{-} & - & \multicolumn{1}{c|}{0.2667} & 0.3111 \\ \hline
\begin{tabular}[c]{@{}l@{}}GPT-3.5 \\ (1-shot)\end{tabular} & \multicolumn{1}{c|}{31.19} & \multicolumn{1}{c|}{35.84} & \multicolumn{1}{c|}{-} & \multicolumn{1}{c|}{-} & - & \multicolumn{1}{c|}{0.5000} & 0.5889 \\ \hline
\end{tabular}

\end{minipage}\hfill 
\begin{minipage}{\columnwidth}
\centering
\small
\caption{Results of Code Generation using "Problem (CWE-aware)" Information.}
\label{tab3}
\setlength{\tabcolsep}{4.0pt}
\renewcommand{\arraystretch}{1.2}

\begin{tabular}{|c|ccccc|cc|}
\hline
\multicolumn{1}{|c|}{\multirow{2}{*}{Model}} & \multicolumn{5}{c|}{CodeBLEU@k} & \multicolumn{2}{c|}{SEC@k} \\ \cline{2-8} 
\multicolumn{1}{|c|}{} & \multicolumn{1}{c|}{k=1} & \multicolumn{1}{c|}{k=2} & \multicolumn{1}{c|}{k=3} & \multicolumn{1}{c|}{k=4} & k=5 & \multicolumn{1}{c|}{k=1} & k=2 \\ \hline
InCoder & \multicolumn{1}{c|}{22.29} & \multicolumn{1}{c|}{27.54} & \multicolumn{1}{c|}{29.43} & \multicolumn{1}{c|}{30.72} & 31.81 & \multicolumn{1}{c|}{0.1000} & 0.1111 \\ \hline
CodeGen & \multicolumn{1}{c|}{23.99} & \multicolumn{1}{c|}{29.14} & \multicolumn{1}{c|}{32.44} & \multicolumn{1}{c|}{34.04} & 35.47 & \multicolumn{1}{c|}{0.3000} & 0.4556 \\ \hline
StarCoder & \multicolumn{1}{c|}{25.97} & \multicolumn{1}{c|}{31.30} & \multicolumn{1}{c|}{34.16} & \multicolumn{1}{c|}{35.68} & 36.32 & \multicolumn{1}{c|} {0.2667} & 0.3556 \\ \hline
GPT-3.5 & \multicolumn{1}{c|}{28.44} & \multicolumn{1}{c|}{33.53} & \multicolumn{1}{c|}{-} & \multicolumn{1}{c|}{-} & - & \multicolumn{1}{c|}{0.7333} & 0.8556 \\ \hline
\end{tabular}

\end{minipage}\hfill 
\end{table*}

\begin{table*}
\begin{minipage}{\columnwidth}

\centering
\small
\caption{Results of Code Repair Performance Evaluation.}
\label{tab4}
\setlength{\tabcolsep}{4.0pt}
\renewcommand{\arraystretch}{1.2}
\begin{tabular}{|c|ccccc|cc|}
\hline
\multirow{2}{*}{Model} & \multicolumn{5}{c|}{CodeBLEU@k} & \multicolumn{2}{c|}{SEC@k} \\ \cline{2-8} 
 & \multicolumn{1}{c|}{k=1} & \multicolumn{1}{c|}{k=2} & \multicolumn{1}{c|}{k=3} & \multicolumn{1}{c|}{k=4} & k=5 & \multicolumn{1}{c|}{k=1} & k=2 \\ \hline
InCoder & \multicolumn{1}{c|}{23.44} & \multicolumn{1}{c|}{33.38} & \multicolumn{1}{c|}{37.53} & \multicolumn{1}{c|}{39.85} & 42.14 & \multicolumn{1}{c|}{0.0556} & 0.0667 \\ \hline
CodeGen & \multicolumn{1}{c|}{18.43} & \multicolumn{1}{c|}{22.31} & \multicolumn{1}{c|}{24.73} & \multicolumn{1}{c|}{25.65} & 26.10 & \multicolumn{1}{c|}{0.0} & 0.0 \\ \hline
StarCoder & \multicolumn{1}{c|}{31.86} & \multicolumn{1}{c|}{38.86} & \multicolumn{1}{c|}{42.09} & \multicolumn{1}{c|}{44.42} & 45.75 & \multicolumn{1}{c|}{0.1000} & 0.1444 \\ \hline
GPT-3.5 & \multicolumn{1}{c|}{25.54} & \multicolumn{1}{c|}{28.47} & \multicolumn{1}{c|}{-} & \multicolumn{1}{c|}{-} & - & \multicolumn{1}{c|}{0.3889} & 0.4556 \\ \hline

\end{tabular}

\end{minipage}\hfill 
\begin{minipage}{\columnwidth}
\small
\setlength{\tabcolsep}{4.0pt}
\renewcommand{\arraystretch}{1.2}
\centering
\caption{Results of Code Repair with "Insecure Code Explanation" Information. }
\label{tab5}

\begin{tabular}{|c|ccccc|cc|}
\hline
\multirow{2}{*}{Model} & \multicolumn{5}{c|}{CodeBLEU@k} & \multicolumn{2}{c|}{SEC@k} \\ \cline{2-8} 
 & \multicolumn{1}{c|}{k=1} & \multicolumn{1}{c|}{k=2} & \multicolumn{1}{c|}{k=3} & \multicolumn{1}{c|}{k=4} & k=5 & \multicolumn{1}{c|}{k=1} & k=2 \\ \hline
InCoder & \multicolumn{1}{c|}{25.23} & \multicolumn{1}{c|}{31.65} & \multicolumn{1}{c|}{36.39} & \multicolumn{1}{c|}{39.12} & 40.28 & \multicolumn{1}{c|}{0.0778} & 0.1111 \\ \hline
CodeGen & \multicolumn{1}{c|}{17.89} & \multicolumn{1}{c|}{21.90} & \multicolumn{1}{c|}{24.72} & \multicolumn{1}{c|}{26.08} & 26.91 & \multicolumn{1}{c|}{0.0} & 0.0 \\ \hline
StarCoder & \multicolumn{1}{c|}{30.41 } & \multicolumn{1}{c|}{39.43} & \multicolumn{1}{c|}{41.95} & \multicolumn{1}{c|}{44.22} & 46.10 & \multicolumn{1}{c|}{0.0778} & 0.1333 \\ \hline
GPT-3.5 & \multicolumn{1}{c|}{25.09} & \multicolumn{1}{c|}{28.07} & \multicolumn{1}{c|}{-} & \multicolumn{1}{c|}{-} & - & \multicolumn{1}{c|}{0.4000} & 0.4778 \\ \hline
\end{tabular}

\end{minipage}\hfill 
\end{table*}

\begin{table*}
\begin{minipage}{\columnwidth}
\footnotesize
\centering
\caption{Detailed results of GPT-3.5 for different vulnerability types in code generation and code repair. Each task shows two experimental settings. "ICE" indicates "Insecure Code Explanation" and CoB@2 is CodeBLEU@2 metric. CWE-Aur indicates the merged type about authentication issue.}
\label{tab6}
\footnotesize
\setlength{\tabcolsep}{2.8pt}
\renewcommand{\arraystretch}{1.3}
\begin{tabular}{|c|cccc||cccc|}
\hline
\multirow{3}{*}{Type} & \multicolumn{4}{c||}{Code Generation (GPT-3.5)} & \multicolumn{4}{c|}{Code Repair (GPT-3.5)} \\ \cline{2-9} 
 & \multicolumn{2}{c|}{"Problem"} & \multicolumn{2}{c||}{\begin{tabular}[c]{@{}c@{}}"Problem \\ (CWE-aware)"\end{tabular}} & \multicolumn{2}{c|}{Without "ICE"} & \multicolumn{2}{c|}{With "ICE"} \\ \cline{2-9} 
 & CoB@2 & \multicolumn{1}{c|}{SEC@2} & CoB@2 & SEC@2 & CoB@2 & \multicolumn{1}{c|}{SEC@2} & CoB@2 & SEC@2 \\ \hline
CWE-787 & 23.21 & \multicolumn{1}{c|}{0.0} & 40.30 & 0.8 & 30.85& \multicolumn{1}{c|}{0.0} & 37.17 & 0.2 \\

CWE-79 & 26.14 & \multicolumn{1}{c|}{0.0} & 27.93 & 0.4 & 23.61& \multicolumn{1}{c|}{0.2} & 26.02 & 0.2 \\

CWE-89 & 22.97 & \multicolumn{1}{c|}{0.0} & 40.96 & 1.0 & 28.83& \multicolumn{1}{c|}{1.0} & 26.30 & 1.0 \\

CWE-78 & 29.59 & \multicolumn{1}{c|}{0.2} & 41.61 & 1.0 & 30.60& \multicolumn{1}{c|}{1.0} & 29.90 & 0.8 \\

CWE-20 & 22.33 & \multicolumn{1}{c|}{0.0} & 35.20 & 1.0 & 29.89& \multicolumn{1}{c|}{0.0} & 30.38 & 0.8 \\

CWE-125 & 23.07 & \multicolumn{1}{c|}{0.0} & 26.85 & 0.8 & 27.35& \multicolumn{1}{c|}{0.0} & 31.91 & 0.0 \\

CWE-22 & 33.32  & \multicolumn{1}{c|}{0.8} & 31.69 & 1.0 & 32.94& \multicolumn{1}{c|}{0.6} & 31.39 & 0.2 \\

CWE-352 & 29.05 & \multicolumn{1}{c|}{0.8} & 31.84 & 0.8 & 35.23& \multicolumn{1}{c|}{0.8} & 28.80 & 1.0 \\

CWE-434 & 37.19 & \multicolumn{1}{c|}{0.8} & 38.67 & 1.0 & 26.93& \multicolumn{1}{c|}{0.0} & 35.63 & 1.0 \\

CWE-Aur & 26.37 & \multicolumn{1}{c|}{0.6} & 39.95 & 0.6 & 28.04& \multicolumn{1}{c|}{0.4} & 25.23 & 0.8 \\

CWE-502 & 19.03 & \multicolumn{1}{c|}{0.2} & 28.20 & 0.8 & 22.32& \multicolumn{1}{c|}{0.0} & 22.51 & 0.0 \\

CWE-77 & 27.98 & \multicolumn{1}{c|}{0.6} & 28.98 & 0.8 & 30.16& \multicolumn{1}{c|}{0.4} & 30.52 & 0.0 \\

CWE-798 & 24.04 & \multicolumn{1}{c|}{0.0} & 29.45 & 0.4 & 30.44& \multicolumn{1}{c|}{0.2} & 27.94 & 0.2 \\

CWE-918 & 28.43  & \multicolumn{1}{c|}{0.0} & 46.45 & 1.0 & 34.96& \multicolumn{1}{c|}{0.0} & 32.01 & 0.4 \\

CWE-362 & 28.08 & \multicolumn{1}{c|}{0.8} & 25.24 & 1.0 & 29.57& \multicolumn{1}{c|}{0.8} & 25.50 & 1.0 \\

CWE-269 & 16.91 & \multicolumn{1}{c|}{0.4} & 35.96 & 1.0 & 18.53& \multicolumn{1}{c|}{0.6} & 23.57 & 0.6 \\

CWE-94 & 23.77 & \multicolumn{1}{c|}{0.0} & 25.23 & 1.0 & 21.81& \multicolumn{1}{c|}{0.6} & 15.00 & 0.2 \\

CWE-276 & 26.47 & \multicolumn{1}{c|}{0.4} & 29.05 & 1.0 & 30.45& \multicolumn{1}{c|}{0.6} & 25.58 & 0.2 \\

Avg & 26.00 & \multicolumn{1}{c|}{0.3111} & 33.53 & 0.8556 & 28.47 & \multicolumn{1}{c|}{0.4556} & 28.07 & 0.4778 \\ \hline
\end{tabular}

\end{minipage}\hfill 
\begin{minipage}{\columnwidth}

\footnotesize
\centering
\caption{Detailed results of Vulnerability Classification Challenges using GPT-3.5, including two sub-tasks: vulnerability detection and vulnerability type prediction. CWE-Aur indicates the merged type about authentication issue.}
\label{tab7}
\renewcommand{\arraystretch}{1.2}
\begin{tabular}{|c|c||cccc|}
\hline
\multirow{2}{*}{Type}  & \begin{tabular}[c]{@{}c@{}}Vulnerability\\ Detection\end{tabular} & \multicolumn{4}{c|}{Vulnerability Type Prediction} \\ \cline{2-6} 
 & accuracy & precision & recall & f1-score & support \\ \hline
CWE-787 & 0.6 & 0.00 & 0.00 & 0.00 & 6 \\
CWE-79 & 0.9 & 0.77 & 1.00 & 0.87 & 10 \\
CWE-89 & 0.8 & 0.67 & 1.00 & 0.80 & 10 \\
CWE-78  & 0.6 & 0.35 & 0.60 & 0.44 & 10\\
CWE-20 & 0.5 & 0.39 & 0.78 & 0.52 & 9 \\
CWE-125 & 0.5 & 0.50 & 0.22 & 0.31 & 9 \\
CWE-22 & 0.8 & 0.83 & 1.00 & 0.91 & 10 \\
CWE-352 & 0.7 & 1.00 & 0.80 & 0.89 & 5 \\
CWE-434 & 0.7 & 0.90 & 0.90 & 0.90 & 10 \\
CWE-Aur & 0.4 & - & - & - & - \\
CWE-502 & 0.9 & 0.78 & 0.78 & 0.78 & 9 \\
CWE-77 & 1.0 & 0.60 & 0.30 & 0.40 & 10 \\
CWE-798 & 0.8 & 0.91 & 1.00 & 0.95 & 10 \\
CWE-918 & 0.9 & 1.00 & 0.60 & 0.75 & 10 \\
CWE-362 & 0.8 & 0.83 & 1.00 & 0.91 & 10 \\
CWE-269 & 0.7 & 0.64 & 0.70 & 0.67 & 10 \\
CWE-94 & 0.7 & 1.00 & 0.50 & 0.67 & 10 \\
CWE-276 & 0.9 & 1.00 & 0.60 & 0.75 & 10 \\ \hline
accuracy & 0.73 & - & - & 0.71 & 158 \\
macro avg & - & 0.72 & 0.69 & 0.68 & 158 \\
weighted avg & - & 0.73 & 0.71 & 0.69 & 158 \\ \hline
\end{tabular}

\end{minipage}\hfill 
\end{table*}

\noindent\textbf{RQ1} \emph{How effective are current large language models in addressing security concerns during code generation, and are certain vulnerability types more likely to be successfully mitigated during code generation?} \\ 
\indent To address RQ1,we evaluate the models' performance in generating code based on clear "Problem" information.  The input prompt used for this experiment is provided in the top left corner of Figure~\ref{exp_prompts}. As shown in Table~\ref{tab2}, among the models except GPT-3.5 (1-shot), StarCoder achieves the highest CodeBLEU@k scores (where k ranges from 2 to 5). However, when considering the SEC@1 or SEC@2 metrics from human evaluation, CodeGen outperforms StarCoder. Interestingly, Incoder achieves comparable CodeBLEU@k results to CodeGen, but its SEC@1 and SEC@2 scores are both 0, indicating that none of its top-2 generated code samples are secure or correct. This can be attributed to Incoder's focus on the filling task due to its causal masking objective during pre-training, affecting its performance in the code generation task. Notably, GPT-3.5 stands out with the best performance in SEC@1, SEC@2, and CodeBLEU@1. It outperforms the second-ranked CodeGen by 59.99\% and 55.56\% in SEC@1 and SEC@2, respectively.

Upon closer examination of GPT-3.5's detailed results of different vulnerability types in this experiment, as shown in the "Problem" column of Table~\ref{tab6}, we observe that GPT-3.5 exhibits varying performance across vulnerability types. For certain types, such as CWE-22, CWE-352, CWE-434, and CWE-362, GPT-3.5 is capable of generating correct insecure code within its top-2 results for most given "Problem". However, for other vulnerability types, GPT-3.5's performance is notably weaker, with most SEC@2 scores being 0.

Overall, the results shed light on the strengths and weaknesses of the evaluated models in addressing security concerns during code generation. GPT-3.5 stands out as the most promising model for generating secure code, but there is room for improvement, especially in handling certain vulnerability types.\\ 


\noindent\textbf{RQ2} \emph{What effective approaches can be devised to improve the security of code generation by large language models, and to what extent can these proposed approaches mitigate security vulnerabilities?} \\ 
\indent Next, we aim to explore ways to enhance the security of code generation by large language models. Intuitively, if the problem can be formulated to also demonstrate potential vulnerabilities, the code LLMs might avoid generating vulnerable code. Thus, we use the "Problem (CWE-aware)" as input to investigate whether incorporating potential vulnerability information can help in generating secure code. The results demonstrate a notable improvement. As shown in Table~\ref{tab3}, the performance of all models increases significantly across both CodeBLEU@k and SEC@k metrics. Notably, using potential vulnerability information leads to substantial gains in the InCoder model, with SEC@1 and SEC@2 increasing from 0 to 0.1 and 0.1111, respectively. Particularly, GPT-3.5 exhibits the most significant improvement, with an increase of 174.95\% and 175.02\% in SEC@1 and SEC@2, respectively.

Similarly, when examining GPT-3.5's detailed results for different vulnerability types in this experiment, as shown in the "Problem (CWE-aware)" column of Table~\ref{tab6}, we make several observations. Firstly, the CodeBLEU@2 scores improve significantly in most cases, particularly for types that performed poorly in the first experiment. Secondly, we find that only 3 of the 18 types have SEC@2 scores lower than 0.8, indicating that incorporating potential vulnerability information greatly enhances the model's ability to generate more secure code. However, we also note that for certain types, such as CWE-79 (Improper Neutralization of Input During Web Page Generation ('Cross-site Scripting')) and CWE-798 (Use of Hard-coded Credentials), although the two scores are improved, they still pose challenges for GPT-3.5 to handle effectively.

In addition to this approach, we aim to investigate whether 1-shot prompting helps in improving the models' performance on our SecuCoGen dataset, i.e., providing a pair of different "Problem" and corresponding "Secure Code" in front of the target "Problem." The results of 1-shot prompting are shown in the last line of Table~\ref{tab2}. We can see that for the description that is "Problem," the results improved across both CodeBLEU@k and SEC@k. Compared with the model using "Problem," the improvements on CodeBLEU@2 and SEC@2 are 37.85\% and 89.30\%, respectively. However, when comparing with the results in Table~\ref{tab3}, 1-shot prompting achieves better results on CodeBLEU@k but performs much worse in SEC@k.\\

\noindent\textbf{RQ3} \emph{How well do existing large language models perform in repairing insecure code?}\\ 
\indent Next, we examine the performance of CodeLLMs in code repair using the SecuCoGen dataset, where the input consists of "Problem" and "Insecure Code." The results of this experiment are shown in Table~\ref{tab4}. Comparing these results with the ones obtained from code generation using the "Problem" input in the first experiment, we find that CodeGen performs significantly worse in the code repair task. All CodeBLEU@k scores decrease, and the SEC@1 and SEC@2 scores both decrease from 0.1667 and 0.2000 to 0. This indicates that CodeGen is not effective at repairing insecure code. In contrast, the other three models, especially InCoder and StarCoder, exhibit higher CodeBLEU@k scores for the repair task compared to code generation. For example, the InCoder model shows improvements of 38.18\% and 44.96\% at CodeBLEU@3 and CodeBLEU@5, respectively, compared to code generation. However, we do not observe significant improvements in the SEC@k metrics for these two models. Moreover, GPT-3.5 improves its performance in both metrics. \\

\noindent\textbf{RQ4} \emph{To what extent does explaining the reasons why the code is insecure help in repairing the code by existing models?}\\ 
\indent We then explore whether including "Insecure Code Explanation" improves the repair of insecure code. The results are shown in Table~\ref{tab7}. Surprisingly, we do not observe significant differences compared to the previous experiments. Most results show slight improvements or declines, indicating that the "Insecure Code Explanation" does not provide significant assistance in this case. We believe there might be two reasons for this. Firstly, the prompt might not be well-designed to instruct the LLMs to effectively repair the vulnerability. Secondly, the tested LLMs might not have strong capabilities in fixing vulnerable code, possibly due to their pre-training tasks or data.

Overall, these results emphasize the complexity of repairing insecure code using existing large language models and underscore the importance of advancing code repair approaches to enhance security in software development practices. Further research and improvements are necessary to develop more effective methods for repairing vulnerable code.\\

\noindent\textbf{RQ5} \emph{Which vulnerability types pose challenges for large language models in vulnerability classification?}\\ 
\indent In vulnerability detection, we collect 5 "Insecure Code" and 5 "Secure Code" samples for each type, and GPT-3.5 performs binary classification to indicate whether the given code is vulnerable or not. As shown in the second column of Table~\ref{tab7}, we find that the overall performance of GPT-3.5 in detecting the presence of vulnerability is relatively good, with an average accuracy of 0.73. Additionally, only 3 types have accuracy scores less than or equal to 0.5, indicating that GPT-3.5 is generally effective in identifying whether code samples are vulnerable.
Next, we test vulnerability type prediction. Here, GPT-3.5 is instructed to identify the type of the given insecure code. Note that we exclude the merged CWE-Aur type in this task to avoid confusing GPT-3.5. The input prompt in this task provides short descriptions of all the types, as shown in the bottom right of Figure~\ref{exp_prompts}. Additionally, we exclude records where GPT-3.5 outputs a type not among the tested 17 types. As shown in the corresponding columns of Table~\ref{tab7}, the overall performance is also relatively good, with an accuracy of 0.71. From the table, we find that for the vulnerability types that perform poorly in the detection task (with accuracy less than or equal to 0.6), they also perform poorly in the multi-class prediction task. These types include CWE-787, CWE-78, CWE-20, and CWE-125.

The results demonstrate that GPT-3.5 has the capability to perform well in both vulnerability detection and vulnerability type prediction tasks, with satisfactory accuracy levels for most vulnerability types. However, certain types, as mentioned above, remain challenging for the model. These findings highlight the potential of large language models in vulnerability classification tasks while also pointing towards areas for further improvement and research.\\

\noindent\textbf{RQ6} \emph{What are the implications of the research findings for the broader software engineering community, and how can developers and researchers leverage large language models more securely in real-world applications?}\\
\indent The research findings presented in this study have several implications for the broader software engineering community and offer insights on leveraging large language models more securely in real-world applications.
\begin{enumerate}[leftmargin=2em]
\item Firstly, our study highlights the potential risks associated with using large language models for code generation, particularly in the context of security vulnerabilities. It emphasizes the importance of considering and mitigating security concerns when employing these models in software development tasks. Developers and researchers should be aware of the risks and limitations of existing models in handling security-related aspects of code generation.

\item Secondly, the SecuCoGen dataset introduced in this paper serves as a valuable resource for evaluating code LLMs from a software security perspective. It provides a curated collection of vulnerable and secure code instances, enabling researchers to benchmark and improve the security-awareness capabilities of code LLMs. The dataset can aid in training more secure and robust models for code generation, repair, and vulnerability classification tasks.

\item Thirdly, the research sheds light on the strengths and weaknesses of current large language models in code generation and repair. Understanding the varying performance of different models across different vulnerability types can guide developers in selecting appropriate models for specific use cases, considering security requirements.

\item Finally, our findings underscore the need for further research and advancements in code repair approaches to enhance security in software engineering practices. As large language models continue to evolve, addressing the challenges of repairing insecure code effectively is crucial for building more trustworthy and secure software systems.
\end{enumerate}

To leverage large language models more securely in real-world applications, developers and researchers should consider:
\begin{itemize}[wide, labelwidth=!, labelindent=3pt]
\item Incorporate Security Awareness: When utilizing large language models for code generation tasks, developers should incorporate potential vulnerability information into input prompts to encourage the models to generate more secure code.

\item Validate Repair Capabilities: Before deploying large language models for code repair tasks, thorough validation of their repair capabilities, especially concerning security vulnerabilities, is essential to avoid introducing new security risks.

\item Dataset Curation: Building comprehensive datasets like SecuCoGen that encompass various vulnerability types and provide clear explanations of insecure code can facilitate the development of more robust and secure models.

\item Continuous Model Improvements: Researchers and developers should continuously work on improving large language models' security-awareness capabilities, addressing the limitations identified in our study and other related research.
\end{itemize}

In conclusion, the findings from this research provide valuable guidance for enhancing the security of large language models in code generation and repair tasks, contributing to the overall improvement of secure software engineering practices. By understanding the implications of these findings, developers and researchers can leverage large language models more securely in real-world applications and mitigate potential security risks associated with code generation tasks.

\section{Related Work}
\subsection{Security Issue of Code LLMs}
The recent advancements in pre-training techniques have led to the development of large-scale pre-trained language models specifically designed for code-related tasks, such as CodeBERT \cite{feng2020codebert}, CodeT5 \cite{wang2021codet5}, PyCodeGPT \cite{CERT}, AlphaCode \cite{li2022competition}, and InCoder \cite{fried2022incoder}. Additionally, Codex \cite{chen2021evaluating} has achieved significant progress with a 12-billion-parameters model capable of addressing 72\% of Python programming problems. However, these models are often trained without explicit consideration of security, as they directly use unsanitized data from open-source repositories like GitHub, which introduces the risk of propagating security vulnerabilities.

Previous studies have analyzed the security concerns of code generated by LLMs. For example, \citet{pearce2022asleep} evaluated the security of GitHub Copilot, which is based on the OpenAI Codex model, and found that Copilot generates insecure code about 40\% of the time. \citet{khoury2023secure} investigated security-relevant coding scenarios and revealed that ChatGPT produced insecure code in 16 cases, with only 7 cases being self-corrected after further prompting. Moreover, experiments conducted by \citet{perry2022users} showed that developers using AI model assistance generated more security vulnerabilities, particularly in string encryption and SQL injection, when interacting with OpenAI's codex-davincii-002 model.

While most research has focused on detecting security issues in code generated by large language models, effective methods for solving these security problems or extensive validation using large and diverse test sets have been limited. Some studies have explored the ability of large language models for zero-shot code repair of vulnerabilities \cite{pearce2023examining, chen2022neural, prenner2022can}. For example, \citet{pearce2023examining} demonstrated that models can generate fixes for security bugs when given a carefully constructed prompt. However, their evaluation also highlighted that the state of the art is not yet sufficient to deliver practical value in a program repair framework. Moreover, these studies have limitations: they focus on a limited number of vulnerability types (e.g., only 7 types in \cite{pearce2023examining}), heavily rely on security tools like CodeQL (which exhibited suboptimal performance in detecting vulnerabilities in our collected "Insecure Code" in SecuCoGen), and utilize relatively simpler problems compared to those found in SecuCoGen \cite{chen2022neural, prenner2022can}.

In comparison with previous works, our study goes beyond vulnerability analysis to include: (1) Creating SecuCoGen, a comprehensive dataset for security evaluation. (2) Proposing effective approaches to enhance code generation security. (3) Investigating code repair and vulnerability classification capabilities. (4) Providing practical implications for secure AI-driven programming.

\subsection{Datasets for secure code generation}
Various datasets have been proposed for code generation, including JuICe \cite{agashe-etal-2019-juice}, CONCODE \cite{iyer-etal-2018-mapping}, DS-1000 \cite{lai2022ds}, and the APPS dataset \cite{apps}. However, these datasets primarily focus on general code generation tasks without a specific emphasis on evaluating the models' ability to generate secure code. Regarding datasets addressing security concerns, most of them are designed for evaluating techniques in vulnerability detection and prediction \cite{arzt2014flowdroid, fan2020ac, nikitopoulos2021crossvul, ponta2019manually}. For code repair tasks, QuixBugs \cite{lin2017quixbugs} consists of programs translated to both Python and Java, each containing a single-line bug. However, this dataset is relatively small, comprising only 40 data instances. Regarding secure code generation, SecurityEval \cite{siddiq2022securityeval} was introduced. It includes 130 Python code samples spanning 75 vulnerability types. Nevertheless, each sample in SecurityEval offers limited information, including an 'ID', a 'Prompt' (a partial source code), and an 'Insecure code' representing a possible vulnerable code example". Moreover, the 'Insecure Code' instances in SecurityEval present relatively simple problems.

Despite the existence of these datasets, there remains a gap in the availability of comprehensive datasets that address security concerns in the context of code generation tasks. To address this gap, we present SecuCoGen, a meticulously curated dataset specifically designed to evaluate large language models' security-awareness in code generation, code repair, and vulnerability classification tasks. SecuCoGen encompasses a diverse set of critical vulnerability types and offers in-depth attributes for each data instance, providing a valuable resource to the research community to advance the state of secure code generation using large language models.

\section{Conclusions And Future Work}

This paper provides a comprehensive study that aims to evaluate and enhance code LLMs from a software security perspective. Extensive experiments on our curated SecuCoGen dataset yield valuable insights into the strengths and limitations of large language models in security-critical software engineering tasks. Our proposed approaches for code generation have demonstrated their effectiveness in enhancing code security and mitigating security vulnerabilities. However, we also identified specific weaknesses in existing LLMs' capabilities, particularly in code repair and vulnerability classification for certain vulnerability types. To advance the field of secure code generation, future research should explore the generalizability of our approaches to other programming languages. Moreover, improving the code repair capabilities of LLMs remains a promising direction, and further research could investigate the effectiveness of integrating domain-specific knowledge and feedback mechanisms to produce more robust and secure code repairs. Overall, this study contributes to a better understanding of LLMs' potential and limitations in addressing security concerns, thereby fostering safer and more reliable software development practices in real-world applications.

\bibliographystyle{ACM-Reference-Format}
\bibliography{sample-base}

\end{document}